\documentclass[twoside,reqno]{qts-proc}
\usepackage{epsfig,cite}
\usepackage{amssymb,amsmath}
\usepackage{times}
\begin{document}
\sloppy \raggedbottom
\setcounter{page}{1}

\newpage
\setcounter{figure}{0}
\setcounter{equation}{0}
\setcounter{footnote}{0}
\setcounter{table}{0}
\setcounter{section}{0}



\title{Study of Dissipative System of Charged Particles by Wigner's Functions}

\runningheads{S.Khademi}{Study of Dissipative systems...}

\begin{start}


\author{S.Khademi}{1},
\coauthor{S.Nasiri}{1,2}, \coauthor{S.Fathi}{1},
\coauthor{F.Taati}{2}

\address{Department of Physics, Zanjan Univ., ZNU, Zanjan, Iran.}{1}
\address{Institute for Advanced Studies in Basic Sciences, IASBS, Zanjan, Iran.}{2}


\begin{Abstract}
Recently, it is shown that the extended phase space formulation of
quantum mechanics is a suitable technique for studying the quantum
dissipative system. Here, as an application of this formalism, we
consider a dissipative system of charged particles interacting
with an external time dependent electric field. Such a system has
been investigating by Buch and Denman, and two solutions with
completely different structure have been obtained for
Schr\"{o}dinger's equation in two different gauges. We demonstrate
how both gauges lead to the same conductivity by generalizing the
gauge transformations to the phase space and using the extended
phase space technique.
\end{Abstract}
\end{start}


\section{Introduction}

Buch and Denman investigated a system of charged particles
interacting with an external time dependent electric field [1].
They obtained two distinct solutions with completely different
physical structures for Schr\"{o}dinger's equation in different
gauges. It seems that, by applying the Extended Phase Space (EPS)
formulation of quantum mechanics, proposed by Sobouti and Nasiri
[2,3], the above discrepancy could be removed. In this respect,
Khademi and Nasiri solved the same problem using EPS method and
obtained an identical result for conductivity as a measurable
quantity in two different gauges [4]. Here we consider a system of
dissipative charged particles interacting with an external time
dependent electric field and study its dynamics using the Wigner
function. It is shown that the expression obtained for the
conductivity is the same as given by Khademi and Nasiri.

The layout of the paper is as follows: In section 2, a brief
review of the EPS formalism is presented. In section 3, by
introducing two distinct gauges the conductivity is calculated
using the Wigner functions in these gauges. Section 4 is devoted
to conclusions.
\section{Review of formulation}

A direct approach to quantum statistical mechanics is proposed by
Sobouti and Nasiri (SN) [2], by extending the conventional phase
space and by applying the canonical quantization procedure to the
extended quantities in this space. Assuming the phase space
coordinates $p$ and $q$ to be independent variables on the virtual
trajectories, allows one to define momenta $\pi_{p}$ and
$\pi_{q}$, conjugate to $p$ and $q$, respectively. This is done by
introducing the extended lagrangian

\begin{equation}
{\cal L}(p,q,{\dot p},{\dot q})=-{\dot p}q-{\dot q}p + {\cal L}^{p}(p,{\dot p})+ {\cal L}^{q}(q,{\dot q}),
\end{equation}
where ${\cal L}^p $ and ${\cal L}^{q}$ are the $p$ and $q$ space
lagrangians of the given system. Using Eq. (1) one may define the
momenta, conjugate to $p$ and $q$, respectively, as follow
\begin{equation}
\pi_{p} = \frac{\partial {\cal  L} }{\partial {\dot p} } = \frac{\partial {\cal L}^p}{\partial {\dot p}}-q,
\end{equation}

\begin{equation}
\pi_{q} = \frac{\partial {\cal  L }}{\partial {\dot q} } = \frac{\partial  {\cal L} ^q}{\partial {\dot q} }-p.
\end{equation}
In the EPS defined by the set of variables $\{p,q, \pi_p,
\pi_q\}$, one may define the extended hamiltonian
\begin{eqnarray}
{\cal H}_{SN}(\pi_p ,\pi_q , p,q) & = & {\dot p}\pi_p +{\dot q}\pi_q -{\cal L}= H(p+\pi_q,q)-H(p,q+\pi_p)\nonumber\\
& = & \sum\frac{1}{n!}\left\{\frac{\partial^nH}{\partial p^n}\pi_{q}^n-\frac{\partial^nH}{\partial q^n}\pi_{p}^n\right\},
\end{eqnarray}
where $H(p,q)$ is the hamiltonian of the system. Using the
canonical quantization rule, the following postulates are
outlined: a) Let $p,q,\pi_p$ and $\pi_q$ be operators in a Hilbert
space, $\bf X$, of all square integrable complex functions,
satisfying the following commutation relations
\begin{equation}
[\pi_{q},q]=-i\hbar,\hspace{2cm}\pi_{q}=-i\hbar\frac{\partial}{\partial q},
\end{equation}

\begin{equation}
[\pi_{p},p]=-i\hbar,\hspace{2cm}\pi_{p}=-i\hbar\frac{\partial}{\partial p},
\end{equation}

\begin{equation}
[p,q]=[\pi_{p},\pi_{q}]=0.
\end{equation}
By virtue of Eqs. (4), the extended hamiltonian, ${\cal H}$, will
also be an operator in ${\bf X}$. b) A state function
$\chi(p,q,t)\in{\bf X}$ is assumed to satisfy the following
dynamical equation
\begin{eqnarray}
i\hbar\frac{\partial \chi}{\partial t}&=&{\cal
H}_{SN}\chi=[H(p-i\hbar\frac{\partial}{\partial q},q)
-H(p,q-i\hbar\frac{\partial}{\partial p})]\chi\nonumber\\
&=&\sum\frac{(-i\hbar)^n}{n!}\left\{\frac{\partial^nH}{\partial
p^n}\frac{\partial^n}{\partial q^n}-\frac{\partial^nH}{\partial
q^n}
\frac{\partial^n}{\partial p^n}\right\}\chi.
\end{eqnarray}
c) The averaging rule for an observable $O(p,q)$, a c-number
operator in this formalism, is given as
\begin{equation}
<O(p,q)>=\int O(p,q)\chi^{*}(p,q,t)dpdq.
\end{equation}
For details of selection procedure of the admissible state
functions, see Sobouti and Nasiri [2].

\section{Dynamics of Dissipative Charged Particle via Wigner's Functions}

In classical level, the Wigner extended hamiltonian is related to
the extended hamiltonian (8) by an extended canonical
transformation. Thus, in quantum level, the Wigner hamiltonian
operator can be obtained by corresponding unitary transformation
as follows [2]
\begin{equation}
{\cal H}_w = U{\cal H}_{SN}U\dagger=H(p+\frac{1}{2}
\pi_q,q-\frac{1}{2}\pi_p)-H(p-\frac{1}{2}\pi_q,q+\frac{1}{2}\pi_p),
\end{equation}
where
\begin{equation}
U=exp(-\frac{i}{2}\pi_q\pi_p)/\hbar.
\end{equation}
The evolution of Wigner's distribution function $w(p,q)$, is given
by Wigner's equation,
\begin{equation}
i \hbar \frac{\partial w}{\partial t} = {\cal H}_ w w.
\end{equation}
The Kanai hamiltonian [5] for a dissipative particle in a medium
with damping constant $\alpha$ is
\begin{equation}
H= \frac{1}{2m} [p-\frac{e}{c} A(q,t)]^2e^{-\alpha t}+ e
\phi(q,t)e^{\alpha t}.
\end{equation}
In Eq. (13), $\phi$ and $A$ are electromagnetic scalar and vector
potentials, respectively.

To study the above problem we choose two distinct gauges, i.e.,
$A$- and $\phi$- gauges in EPS. The generalization of the gauge
transformations to the EPS is done by Khademi and Nasiri [4]. They
have shown that, the gauge function in EPS becomes a function of
ordinary phase space coordinates $p$ and $q$ which is the sum of
the gauge functions in $q$ and $p$ spaces. The corresponding
extended unitary transformation which gives the extended
hamiltonian operator in different gauges is the product of the
unitary transformations in $q$ and $p$ spaces.

\subsection{$A$-gauge} A time dependent uniform electric field may
be generated by setting
\begin{equation}
A(t)=-c \int^t \exp(\alpha \lambda) E(\lambda) d\lambda.
\end{equation}
Note that $A(t)$ depends only on time. In this gauge Wigner's
extended hamiltonian assumes the following form
\begin{equation}
{\cal H}^A _w =\frac {1}{m} p \exp(-\alpha t)\pi_q - \frac
{e}{mc}A(t)\exp(-\alpha t)\pi_q.
\end{equation}
Using this hamiltonian in Eq. (12), one gets
\begin{equation}
i \hbar \frac{\partial w}{\partial t} = {\cal H}^A_w w =-i
 \frac{\hbar}{m} \exp(-\alpha t)p \frac{\partial w}{\partial q}+
\frac{i \hbar}{m} \frac{e}{c} \exp(- \alpha t)A(t) \frac{\partial
w}{\partial q}.
\end{equation}
Taking the transformation
\begin{equation}
\xi=\frac{-eE(t)}{imw(\alpha+iw)} + q,
\end{equation}
\begin{equation}
\eta=p,
\end{equation}
\begin{equation}
\tau=t,
\end{equation}
and using them in Eq. (16), one gets
\begin{equation}
\frac {-\eta}{m} e^{-\alpha \tau} \frac{\partial w}{\partial \xi}
= \frac {\partial w}{\partial \tau}.
\end{equation}
Equation (16) can be solved and yields
\begin{equation}
w=c \exp(\frac{-k}{\alpha} e^{-\alpha t}) \exp(\frac{-km}{a}\xi)
\delta(\eta - a),
\end{equation}
where $c$ and $k$ are normalization and separation constants. The
conductivity for a system of $N$ particles is given by
\begin{equation}
\sigma = \frac{Ne<\dot q>}{E(t)},
\end{equation}
where
\begin{equation}
\dot q = \frac{\partial {\cal H}^A_w}{\partial
\pi_q}=\frac{\eta}{m} \exp(-\alpha \tau) +
\frac{eE(\tau)}{m(\alpha+iw)}.
\end{equation}
Then
\begin{equation}
\sigma = \frac{Ne^2}{m(\alpha+iw)}
\end{equation}
Note that the first term in Eq. (23) is a transient term and
vanishes as $t\rightarrow \infty$.

\subsection{$\phi$-gauge} In $\phi$-gauge, $\phi(q,t)$ is define as

\begin{equation}
\phi(q,t)=-qE(t).
\end{equation}
The Wigner extended hamiltonian becomes
\begin{equation}
{\cal H}^\phi_w = \frac{1}{m} \exp(-\alpha t) p \phi_q +
eE(t)\exp(\alpha t) \phi_p.
\end{equation}
The evolution equation for $w$ gives

\begin{equation}
i\hbar \frac{\partial w}{\partial t} ={\cal  H}^\phi_w w =
\frac{-i\hbar}{m} \exp(-\alpha t) p \frac{\partial w}{\partial q}
- i\hbar eE(t) \exp(\alpha t) \frac{\partial w}{\partial p}.
\end{equation}
Taking the transformation
\begin{equation}
\xi^{\prime} = q - \frac{eE(t)}{imw(\alpha+iw)},
\end{equation}
\begin{equation}
\eta^{\prime} = p - \frac{eE(t) \exp(\alpha t)}{\alpha+iw},
\end{equation}
\begin{equation}
\tau^{\prime}  = t.
\end{equation}
one has

\begin{equation}
\frac{-\eta^{\prime}}{m} \exp(-\alpha t) \frac{\partial
w}{\partial \xi^{\prime}} =\frac{\partial w}{\partial
\tau^{\prime}}
\end{equation}
which has the same form as Eq. (20). THus the Wigner function has
the same mathematical structure in both gauges. To calculate the
conductivity one may uses
\begin{equation}
\dot q = \frac{\partial {\cal H}^\phi_w}{\partial
\pi_q}=\frac{\eta^{\prime}}{m} \exp(-\alpha t) +
\frac{eE(\tau^{\prime})}{m(\alpha+iw)}.
\end{equation}
One may easily show that $\dot q(\xi,\eta,\tau)$ and $\dot
q(\xi^{\prime},\eta^{\prime},\tau^{\prime})$ have exactly the same
functional forms. Therefore, the solution of Eqs. (16) and (27)
will be identical, giving the same physical results for
conductivity in $A$ and $\phi$ gauges.


\section{Conclusions}

In an earlier work, Khademi and Nasiri [4], used the EPS
formulation proposed by Sobouti and Nasiri [2], to study the
quantum behavior of a system of dissipative charged particles in a
time dependent electric field and, in contrast to the Bush and
Dennman, they showed that the conductivity for the system is gauge
independent. In this paper, using the Wigner functions, we studied
the same problem and obtained the same expression for the
conductivity in two different $A$ and $\phi$ gauges. The results
are the same as those obtained by Khademi and Nasiri. This is
because, the Wigner hamiltonian can be obtained by an extended
canonical transformation on the SN hamiltonian [2]. Then, one
expects that in quantum level the corresponding quantum
distribution functions to be related by a unitary transformation.
Therefore, the physical quantities, i.e., the conductivity, must
be the same in two approaches.

%


\section*{Acknowledgments}

We are grateful to Prof. Sobouti for his helpful comments.


\end{document}